 \documentstyle[11pt,aasms4,flushrt,epsf,rotate]{article}

\def\bt{\begin{tabbing}}
\def\et{\end{tabbing}}
\def\beq#1{\begin{equation}\label{#1}}
\def\eeq{\end{equation}}

\def\xbf{{\bf x}}
\def\xrm{{\rm x}}

\def\ybf{{\bf y}}
\def\nbf{{\bf n}}

\def\Fbf{{\bf F}}

\def\Cbf{{\bf C}}
\def\alphaHat{\hat{\alpha}}

\def\syn{synchrotron~}
\def\syyn{synchrotron}

\def\microk{$\mu$K~}
\def\mmicrok{$\mu$K}
\def\microm{$\mu$m~}

\def\mj{MJy/sr}
\def\letter{$Letter$~}

\def\From{From~}
\def\rms{{\it rms~}}

\def\spose#1{\hbox to 0pt{#1\hss}}
\def\simlt{\mathrel{\spose{\lower 3pt\hbox{$\mathchar"218$}}
     \raise 2.0pt\hbox{$\mathchar"13C$}}}
\def\simgt{\mathrel{\spose{\lower 3pt\hbox{$\mathchar"218$}}
     \raise 2.0pt\hbox{$\mathchar"13E$}}}
\def\simpropto{\mathrel{\spose{\lower 3pt\hbox{$\mathchar"218$}}
     \raise 2.0pt\hbox{$\propto$}}}

\def\pp{\noindent\parshape 2 0truecm 13.6truecm 1truecm 12.6truecm}
\def\rn{\pp}

\def\bfig{\begin{figure}[h] \centerline{\hbox{}}\vfill}
\def\efig{\end{figure}\vfill\newpage}

 \journalid{337}{15 January 1989}
 \articleid{11}{14}

 \slugcomment{{\it{ApJ Lett}}, 482:L17-20 (1997)}

\begin{document}

\title{GALACTIC MICROWAVE EMISSION AT DEGREE ANGULAR SCALES}

\author{Ang\'elica de Oliveira-Costa$^{1,2}$}
\author{A. Kogut$^3$}
\author{Mark J. Devlin$^4$}
\author{C. Barth Netterfield$^5$}
\author{Lyman A. Page$^1$}
\author{Edward J. Wollack$^6$}
\affil {$^1$Princeton University, Department of Physics, Jadwin Hall, 
       Princeton, NJ 08544; angelica@tatania.princeton.edu}
\affil {$^2$Institute for Advanced Study, Olden Lane, Princeton, 
       NJ 08540}
\affil {$^3$Hughes STX Corporation, Laboratory for Astronomy and Solar
       Physics, Code 685, NASA/GSFC, Greenbelt, MD 20771}
\affil {$^4$University of Pennsylvania, Department of Physics and Astronomy,
       David Rittenhouse Laboratory, Philadelphia, PA 19104}
\affil {$^5$California Institute of Technology, MS 59-33, Pasadena, CA 91125}
\affil {$^6$National Radio Astronomy Observatory, 
       Central Development Laboratory, 2015 Ivy Road, 
       Charlottesville, VA 22903}

\begin{abstract}
We cross-correlate the Saskatoon Ka and Q-Band Cosmic Microwave 
Background (CMB) data with different maps to quantify possible 
foreground contamination. 
We detect a marginal correlation ($\simgt$ 2$\sigma$) with the Diffuse 
Infrared Background Experiment (DIRBE) 240, 140 and 100 \microm maps, 
but we find no significant correlation with point sources, with the 
Haslam 408~MHz map or with the Reich and Reich 1420~MHz map. 
The \rms amplitude of the component correlated with DIRBE is about 
20\% of the CMB signal. Interpreting this component as free-free emission, 
this normalization agrees with that of Kogut et al. (1996a; 1996b) and 
supports the hypothesis that the spatial correlation between dust and 
warm ionized gas observed on large angular scales persists to smaller 
angular scales. Subtracting this contribution from the CMB data reduces 
the normalization of the Saskatoon power spectrum by only a few percent. 
\end{abstract}


% That's it for the front matter.  On to the main body of the paper.
% We'll only put in tutorial remarks at the beginning of each section
% so you can see entire sections together.
%
% In the first two sections, you should notice the use of the LaTeX \cite
% command to identify citations.  The citations are tied to the
% reference list via symbolic tags.  We have chosen the first three
% characters of the first author's name plus the last two numeral of the
% year of publication.  The corresponding reference has a \bibitem
% command in the reference list below.
%
% Please go to the LaTeX manual for a complete description of the
% \cite-\bibitem mechanism.

%%%%%%%%%%%%%%%%%%% TEXT: %%%%%%%%%%%%%%%%%%%%%%%%%%%%%%%

\section{INTRODUCTION}
 
One of the major concerns in any Cosmic Microwave Background (CMB) 
anisotropy analysis is to determine if the observed signal is due to 
real CMB fluctuations or due to some foreground contaminant. 
At the frequency range and angular scale of the Saskatoon experiment 
(Wollack et al. 1997, hereafter W97; Netterfield et al. 1997, 
hereafter N97), there are two major potential sources of 
foreground contamination: diffuse Galactic emission and unresolved 
point sources.

The diffuse Galactic contamination includes three components: 
\syn and free-free radiation, which are important mainly at 
frequencies below 60~GHz, and thermal emission from dust particles, 
which is important mainly at frequencies above 60~GHz 
(see, e.g., Partridge 1995; Bennett et al. 1992, Brandt et al. 1994, 
Bennett et al. 1996). 
\From the theoretical point of view, it is possible to distinguish 
these three components by observing their different frequency 
dependence and spatial morphology. In practice, however, there is 
no emission component for which both the frequency dependence and 
spatial template are currently well known (see, e.g., Kogut et al. 
1996a, hereafter K96a, and references therein). 

Upper limits have been placed on the different contaminants at the 
frequency range and angular scale of the Saskatoon experiment.
For instance, by extrapolating radio maps at 408~MHz 
(Haslam et al. 1981) and 1420~MHz (Reich and Reich 1988) to 40~GHz,
N97 place an upper limit on \rms fluctuations in the \syn 
emission of 3 \microk within the Saskatoon observing region. 
\From analysis done on H$\alpha$ maps of the Saskatoon observing 
region, Gaustad et al. (1996) and Simonetti et al. (1996) place an 
upper limit on \rms fluctuations in the diffuse free-free 
emission of 2~\microk at 40 GHz. An analysis of the Galactic emission in 
the Differential Microwave Radiometer (DMR) map at 53~GHz places limits 
of (3.4$\pm$3.7)~\microk for the \syn emission, (7.1$\pm$1.7)~\microk 
for the free-free, and (2.7$\pm$1.3)~\microk for the dust emission, 
for $|b|>$30$^{\circ}$ (Kogut et al. 1996b, hereafter K96b). 
Extrapolating these DMR results to degree angular scales, Tegmark and  
Efstathiou (1996) obtained an upper limit of 2~\microk
for dust emission at 40~GHz.

The purpose of this \letter is to use the Saskatoon data (W97, N97; 
Tegmark et al. 1997, hereafter T97) to estimate the Galactic 
emission at degree angular scales. 
In the following sections, we present the technique used to identify 
the Galactic emission and the results obtained from the cross-correlation
of the Saskatoon data with different Galactic templates. 

\section{THE METHOD}

The Saskatoon data set consists of $N$=2586 observed 
data points y$_{SK}^i$ (not including ``ring data''), each containing 
the true CMB temperature fluctuations in the sky x$_{CMB}$ convolved 
with some beam function, and with noise n$_i$ added afterwards. 
We assume that the Saskatoon data are a superposition of the CMB 
fluctuations and Galactic signals whose angular distributions are 
traced by an external data set (K96a; K96b). Combining these 
numbers into vectors $\ybf_{SK}$, $\xbf_{CMB}$ and $\nbf$, 
respectively, we can therefore write
\beq{signals}
  \ybf_{SK} ~=~ \Fbf \xbf_{CMB} ~+~ \nbf 
  	    ~+~ {\alpha} \Fbf \xbf_{Gal},
\eeq
where $\Fbf$ is the beam function matrix defined in T97, x$_{Gal}^i$
are the brightness fluctuations of the Galactic template map 
(not necessarily in temperature units), and $\alpha$ is the 
coefficient that converts the Galactic template into antenna 
temperature.
Since $\langle \xbf_{CMB} \rangle  = \langle \nbf \rangle  = $0, and 
      $\Fbf \xbf_{Gal}$ ($\equiv \ybf_{Gal}$)
is a constant vector, the data covariance matrix is given by
\beq{varCMB}
  \Cbf ~\equiv~ 
       \langle  \ybf_{SK} \ybf_{SK}^T \rangle  - 
       \langle  \ybf_{SK} \rangle  \langle \ybf_{SK}^T \rangle  ~=~
       \Fbf \langle  \xbf_{CMB} \xbf_{CMB}^T \rangle  \Fbf^T + 
       \langle  \nbf \nbf^T \rangle,
\eeq
where $\langle \xbf_{CMB} \xbf_{CMB}^T \rangle $ is the correlation between 
the data points and themselves, and $\langle \nbf \nbf^T \rangle $ is the 
noise covariance matrix. The former is given by the angular power 
spectrum $C_{\ell}$ through the familiar relation
\beq{corrfunc}
    \langle  \xrm_{CMB}^i \xrm_{CMB}^j \rangle  ~=~ 
    \sum_{\ell =0}^{\infty}  
      \frac{2 \ell + 1}{4 \pi}  P_{\ell}(\cos \theta_{i,j}) 
      C_{\ell},
\eeq
where $\ell$ is the multipole number, $P_{\ell}$ are the Legendre 
polynomials, and $\theta_{i,j}$ is the angle between pixels $i$ and $j$. 

By minimizing
$  \chi^2 \equiv 
          (\ybf_{SK} - {\alpha} \ybf_{Gal})^T 
          {\bf C}^{-1}
          (\ybf_{SK} - {\alpha} \ybf_{Gal})
$
we can obtain $\alphaHat$, the minimum-variance estimate 
of $\alpha$. 
\From $\frac{\partial \chi^2}{\partial \alpha} \equiv 0$, we
find that 
\beq{alpha}
   \alphaHat ~ = ~ 
   \frac{\ybf_{Gal}^T {\bf C}^{-1} \ybf_{SK}}
	{\ybf_{Gal}^T {\bf C}^{-1} \ybf_{Gal}}
\eeq
with variance 
\beq{varalpha}
   \sigma_{\alphaHat}^2 ~ = ~ 
   \frac{1}{(\ybf_{Gal}^T {\bf C}^{-1} \ybf_{Gal})}.
\eeq
This variance accounts for the effect of chance alignments
between the CMB and the various template maps, since the CMB 
anisotropy term is included in equation~(\ref{corrfunc}). 
We compute $\Cbf$ as described in T97, with a flat power spectrum
$C_{\ell}$ = 6$Q^2/\ell (\ell + 1)$ normalized to $Q$=47~\microk.
Note that our analysis is performed on the Saskatoon scan data
$\ybf_{SK}$, and not on the Wiener-filtered Saskatoon maps. 
 
\section{DATA ANALYSIS AND RESULTS}

The analysis is based on the 1993-1995 data from Saskatoon 
experiment (W97, N97, T97). We treated the Ka-Band data 
(26 to 36~GHz) separately from the Q-Band data (36-46~GHz) 
in order to gain additional frequency information on any 
correlated emission. However, it is important to remember that 
observations made on Ka-Band constitute less than a quarter
of the total data set, and therefore give substantially noisier
estimates. 

The Saskatoon data are insensitive to the monopole ($\ell$=0), 
and only marginally sensitive to other low order $\ell$.  
Accordingly, when we convolve the template maps with the Saskatoon 
beam function, we are removing the mean of the templates, as well as
large angular scale structures. 
As a consequence, our results depend predominantly on the small 
scale intensity variations in the templates ($\ell$$>$30) and are 
insensitive to the zero levels of the Saskatoon data and the 
template maps. 

We cross-correlate the Saskatoon data with two different \syn 
templates: the 408~MHz survey (Haslam et al. 1981) and
the 1420~MHz survey (Reich and Reich 1988).
To study dust and free-free emission, we cross-correlate the 
Saskatoon data with three Diffuse Infrared Background Experiment 
(DIRBE) sky maps at wavelengths 240, 140 and 100~\microm (Boggess et
al. 1992). In order to study the extent of point source 
contamination in the Saskatoon data, we cross-correlate 
with the 1~Jy catalog of point sources at 5~GHz (K\"uhr et al. 1981). 
The templates used in this analysis, as well as the previously 
described Saskatoon data, are shown in Figure~\ref{maps}. 

Tables~\ref{tab:tabCorrKa} and~\ref{tab:tabCorrQ} show the 
coefficients $\alphaHat$ and the resulting fluctuations in antenna 
temperature in the Saskatoon Ka and Q-Band data, 
$\Delta T = \alphaHat~\sigma_{Gal}$, where $\sigma_{Gal}$ is the standard 
deviation of the template map, 
\beq{sigmagal}
  \sigma_{Gal}^2~=~\frac{\ybf_{Gal}^T~\ybf_{Gal}}{N}.
\eeq
The values of $\sigma_{Gal}$ were obtained directly from the template
maps after removing monopole and dipole, and smoothing them by 
a 1$^{\circ}$ Gaussian, while $\sigma_{Gal}$ for the Saskatoon data 
correspond to the \rms variance at 1$^{\circ}$, defined as 
\beq{sigmaSask}
  \sigma_{SK}^2 = \sum_{\ell=0}^{\infty} 
                \left( \frac{2\ell+1}{4\pi} \right) 
		C_{\ell} W_{\ell}^2,
\eeq
where $W_{\ell}$ is the Saskatoon beam function as defined in N97.
The \syn templates, as well as the point source template, are found 
to be uncorrelated with the Saskatoon 
data\footnote{
We remind the reader that systematic effects (such as striping) are 
present in the \syn radio maps, and may be partially responsible 
for the null results of the \syn cross-correlations.}. 
All three DIRBE far-infrared templates show a correlation, indicating 
a detection of signal with common spatial structure in the two data sets.
Since the three DIRBE maps trace the same Galactic contamination 
component, they  provide nearly identical estimates of the Galactic 
emission. 
The error bars on the correlation between any of the DIRBE 
templates and the Saskatoon map are dominated by noise and CMB 
signal (random alignments) in the Saskatoon map. 
For definiteness, we use the DIRBE 100 \microm channel and the
correlations obtained for the Saskatoon Q-Band data when placing 
limits below, since these are the least noisy channels.

K96 detect a positive correlation between the DIRBE far-infrared 
maps and the DMR maps at 31.5, 53, and 90~GHz. \From the spectral 
index of the correlation (rising strongly at long wavelengths)
they identify the source as a superposition of dust and 
free-free emission.
The weaker signal and more restricted frequency coverage preclude 
such an unambiguous identification in the Saskatoon data set.
The spectral index (of the emission component correlated with 
DIRBE) between the Ka and Q-band, 
  	$ \beta = - 0.4 \pm 2.1$,  % For 100 \microm
is compatible with an origin from dust emission, free-free emission, 
or chance alignments. However, the amplitude of the signal is much 
larger than expected for dust 
emission\footnote{ 
If the entire Saskatoon-DIRBE correlation is due to dust emission, 
then DMR should see a larger signal.},
leaving free-free emission and chance alignments as the alternatives. 
As mentioned, our calculation of $\sigma_{\alphaHat}$ includes the 
effect of chance alignments between the CMB and the various template maps. 
Assuming that the  hypothesis of K96 can be extended to Saskatoon 
scales, we argue that the correlation between the DIRBE template and 
the Saskatoon data is most likely due to free-free contamination.  
As shown in Table~\ref{tab:tabCorrQ}, the probability that such a strong 
correlation is caused by chance alignments is a few percent.

We tested the cross-correlation technique by analyzing 1000 constrained 
realizations of CMB and Saskatoon instrument noise.
We recovered unbiased estimates $\alphaHat$ with a variance in excellent 
agreement with equation (\ref{varalpha}). 
As an additional test, we generated simulated DIRBE patches by 
replacing our DIRBE region with an equivalent patch selected from 
somewhere else in the sky. 
Since the \rms signal in the DIRBE map depends strongly on latitude, 
we restricted the patches to lie in the latitude range 
20$^{\circ}$ $< \mid b \mid <$ 34$^{\circ}$, corresponding to the latitude 
range of the Saskatoon observing region. In Figure~\ref{histograms}, 
we plot the cumulative distribution for the correlation 
  ${\rm C} \equiv \left( \frac{\alphaHat}{\sigma_{Gal}} \right) 
           \sigma_{Gal}^{\rm NCP}$ 
between Saskatoon Q-Band data and the 288 selected patches of the 
DIRBE 100 \microk map, where $\sigma_{Gal}^{\rm NCP}$ is the standard
deviation of the patch actually observed by Saskatoon. 
This distribution is seen to agree well with a Gaussian distribution with 
variance given by equation (\ref{varalpha}), even though the statistical 
properties of the DIRBE patches themselves are, of course, not Gaussian. 
Specifically, 284 of the 288 patches (or 98.6\% of them) are less 
correlated with the the Saskatoon Q-Band data than the correct DIRBE 
patch, in good agreement with the significance level of 97\% computed 
using a Gaussian distribution. The same tests were carried out for the 
DIRBE 240 and 140 \microm maps and for the Saskatoon Ka-Band data, 
giving similar results.

Due to the proximity of the Saskatoon observing region to the
Galactic plane, one might conjecture that the bulk of the 
Saskatoon-DIRBE correlation is not due to a small scale physical 
association of dust and free-free emission, but is caused by 
large scale gradients common to all three diffuse Galactic 
foreground components (including \syyn). However, the data do not
support this hypothesis: we cross-correlated the DIRBE 100
\microm and Haslam maps of the Saskatoon observing region and 
found no significant correlation. 

In summary, the Saskatoon-DIRBE correlation at 30 and 40 GHz is 
compatible with the free-free emission detected by DMR at 31.5 GHz,
assuming an $\ell^{-3}$ power spectrum for diffuse free-free emission.
The Saskatoon value is slightly larger than upper limits from direct 
H$\alpha$ images, although the uncertainties are substantial.
We note that the inferred H$\alpha$ limits assume an electron temperature
$T_e \approx 8000$ K; a lower electron temperature would produce a 
larger microwave signal for fixed H$\alpha$ intensity.
A compilation of free-free limits and detections on various angular 
scales is given in Table~\ref{tab:tabff}. 

\section{CONCLUSIONS}

The cross-correlation technique introduced by K96a and K96b 
provides a powerful way of measuring the level of Galactic 
foreground contamination.

The \syn templates, as well as the point source template, 
are uncorrelated with the Saskatoon data. We find a 
% marginally significant 
cross-correlation (at 97\% confidence) between the 
Saskatoon Q-Band data and the DIRBE 100 \microm map. 
The \rms amplitude of the contamination correlated with DIRBE
100 \microm is $\approx$ 17 \microk at 40 GHz.
We argue that the hypothesis of free-free contamination at degree 
angular scales is the most likely explanation for this correlated 
emission. Accordingly, the spatial correlation between dust and 
warm ionized gas observed on large angular scales seems to persist 
down to the smaller angular scales.

As reported by N97, the angular power spectrum from the Saskatoon 
data is $\delta T_{\ell}$=49$^{+8}_{-5}$ \microk at $l$=87
(corresponding to \rms fluctuations around 90 \microk on degree scales), 
which is a much higher signal than any of the contributions from the 
foreground contaminants discussed above, confirming that the Saskatoon
data is not seriously contaminated by foreground sources. 
Since the foreground and the CMB signals approximately add in quadrature, 
a foreground signal with 20\% of the CMB \rms causes the CMB
fluctuations to be over-estimated by $\sqrt{1+0.20^2} - 1 \approx 2\%$. 

\bigskip
\bigskip

In the course of this analysis, we became aware that Erik Leitch 
and collaborators have found a correlation between IRAS 100 \microm 
map and Owens Valley data; these results are forthcoming. 
We would like to thank Norman C. Jarosik, Max Tegmark and David T.
Wilkinson for many useful comments on the manuscript. 
Support for this work was provided by a David \& Lucile Packard 
Foundation Fellowship. 
AC acknowledges the RAPT Foundation for financial support under 
process No. 314159(IIH). 
We acknowledge the NASA office of Space Sciences, the COBE flight 
team, and all those who helped process and analyze the DIRBE data.

%%%%%%%%%%%%%%%%%%%%%% TABLES: %%%%%%%%%%%%%%%%%%%%%%%%%%%%%

\clearpage

\begin{table}
\caption{Correlations with the Saskatoon (SK) Ka-Band (30~GHz)$^{(a)}$}
\begin{center}      
\begin{tabular}{lrcccrr}
\hline
\hline
\multicolumn{1}{l}{Template}           	                    &
\multicolumn{1}{c}{$\alphaHat\pm\sigma_{\alphaHat}^{(b)}$}  & 
\multicolumn{1}{c}{$\frac{\alphaHat}{\sigma_{\alphaHat}}$}  & 
\multicolumn{1}{c}{Significance$^{(c)}$}                    &
%%%%%%%%%%     {c}{$erf\left(\frac{\alpha}{\sqrt{2}\sigma_{\alpha}}\right)$}&
\multicolumn{1}{c}{$\sigma_{Gal}^{(d)}$} 	      & 
\multicolumn{1}{c}{$\Delta T \equiv \alphaHat \sigma_{Gal}$}&
\multicolumn{1}{r}{$ \frac{\Delta T}{\sigma_{SK}}^{(e)}$}  \\
  & & 
  & \multicolumn{1}{c}{(\%)} 
& & \multicolumn{1}{c}{(\mmicrok)} 
  & \multicolumn{1}{r}{(\%)} \\
\hline
 SK (Ka-Band)	   &   1.00$\pm$0.04 &23.4 &100 &90.9  &   90.9$\pm$3.9   & 100  \\
 DIRBE 100 \microm &   11.8$\pm$10.4 & 1.1 &87  & 1.30 &   15.3$\pm$13.5  &  17  \\
 Point Sources     &    4.3$\pm$10.7 & 0.4 &66  & 0.50 &    2.2$\pm$5.3   &   2  \\
 408~MHz survey	   & $-$4.3$\pm$9.6  &     &    & 1.33 & $-$5.8$\pm$12.9  &$-$6  \\
 1420~MHz survey   &$-$0.04$\pm$0.31 &     &    & 0.03 &$-$0.00$\pm$0.01  &$-$0.002\\
\hline
\end{tabular}
\end{center}
$^{(a)}$ 30~GHz and 40~GHz are, respectively, the Saskatoon Ka and Q-Band 
centers for a flat spectrum in antenna temperature. \\
$^{(b)}$ For the Saskatoon Ka and Q-Band data, $\alphaHat$ has no 
units since these templates have units \mmicrok. 
For the DIRBE template, $\alphaHat$ has units \microk (\mj)$^{-1}$~
since this template has units \mj.
For the 408 MHz and 1420 MHz surveys, $\alphaHat$ has units 
\microk K$^{-1}$ since these templates have units K. \\
$^{(c)}$ The probability that a Gaussian random variable with mean zero
and standard deviation $\sigma_{\alphaHat}$ does not exceed $\alphaHat$.\\
$^{(d)}$ The Saskatoon Ka and Q-Band data, 
as well as the point source template,
have units \mmicrok, the DIRBE far-infrared data has units \mj, and
the two \syn templates have units K. \\
$^{(e)}$ The correlation between the Saskatoon data and the different
templates is given as fluctuation percentage, defined as 
$\left( \frac{\alphaHat}{\sigma_{SK}} \right) \sigma_{Gal}$,  
where $\sigma_{SK}$=90.9 \microk is the \rms CMB fluctuations at
1$^{\circ}$.
% %%
% $^{(e)}$ $\alphaHat$ has no units  
% since the template has units \microk. \\
% %%
% $^{(f)}$ $\alphaHat$ has units \microk (\mj)$^{-1}$~
% since the template has units \mj. \\
% %%
% $^{(g)}$ $\alphaHat$ has units \microk K$^{-1}$
% since the template has units K. \\
% %%
\label{tab:tabCorrKa}
\end{table}

\begin{table}
\caption{Correlations with the Saskatoon (SK) Q-Band (40~GHz)$^{(a)}$}
\begin{center}      
\begin{tabular}{lrcccrr}
\hline
\hline
\multicolumn{1}{l}{Template}           	                    &
\multicolumn{1}{c}{$\alphaHat\pm\sigma_{\alphaHat}^{(b)}$}  & 
\multicolumn{1}{c}{$\frac{\alphaHat}{\sigma_{\alphaHat}}$}  & 
\multicolumn{1}{c}{Significance$^{(c)}$}                    &
%%%%%%%%%%     {c}{$erf\left(\frac{\alpha}{\sqrt{2}\sigma_{\alpha}}\right)$}&
\multicolumn{1}{c}{$\sigma_{Gal}^{(d)}$} 	      & 
\multicolumn{1}{c}{$\Delta T \equiv \alphaHat \sigma_{Gal}$}&
\multicolumn{1}{r}{$ \frac{\Delta T}{\sigma_{SK}}^{(e)}$}  \\
  & & 
  & \multicolumn{1}{c}{(\%)} 
& & \multicolumn{1}{c}{(\mmicrok)} 
  & \multicolumn{1}{r}{(\%)} \\
\hline
 SK (Q-Band)	   &   1.00$\pm$0.02 &48.3 &100  & 90.9 &   90.9$\pm$1.9  &  100 \\
 DIRBE 100 \microm &   15.0$\pm$8.1  & 1.9 &97   & 1.30 &   19.5$\pm$10.5 &   21 \\
 Point Sources     &    0.2$\pm$3.5  & 0.1 &52   & 0.50 &    0.1$\pm$1.7  &  0.1 \\
 408~MHz survey	   & $-$4.6$\pm$6.8  &     &     & 1.34 & $-$6.1$\pm$9.1  &$-$7  \\
 1420~MHz survey   & $-$0.1$\pm$0.2  &     &     & 0.03 &$-$0.00$\pm$0.01 &$-$0.003\\
\hline
\end{tabular}
\end{center}
\label{tab:tabCorrQ}
\end{table}

\begin{table}
\caption{Free-Free observations scaled to 40~GHz}
\begin{center}     
\begin{tabular}{lccr}
\hline
\hline
\multicolumn{1}{l}{References}           	&
\multicolumn{1}{c}{Angular Resolution}          &
\multicolumn{1}{c}{Patch Size}                  &
\multicolumn{1}{r}{$\Delta T$} 	\\
  & & &(\mmicrok) \\
\hline
Fomalont et al. 1993   &1  $ \arcmin$ &7  $\arcmin \times$7  $ \arcmin$	&$<$2.4 \\
Gaustad et al. 1996    &0.1$^{\circ}$ &7  $^{\circ}\times$7  $^{\circ}$	&$<$2.2 \\
Reynolds et al. 1992   &0.8$^{\circ}$ &12 $^{\circ}\times$10 $^{\circ}$	&   2.6 \\
Simonetti et al. 1996  &1  $^{\circ}$ &7.5$^{\circ}\times$7.5$^{\circ}$	&$<$2.1 \\
This work$^{(a)}$      &1  $^{\circ}$ &7.5$^{\circ}\times$7.5$^{\circ}$	&17.5$\pm$ 9.5 \\
K96a      	       &10 $^{\circ}$ &$|b|>$30$^{\circ}$		&10.0$\pm$ 3.9 \\
K96b                   &10 $^{\circ}$ &$|b|>$30$^{\circ}$		&15.3$\pm$ 3.6 \\
\hline
\end{tabular}
\end{center}
$^{(a)}$ Assuming dust emission with $\beta$=2, about 10\% of
the observed correlation between Saskatoon Q-Band data and DIRBE
100 \microm template is due to dust and 90\% is due to free-free emission. 
Adjusting the value (19.5$\pm$10.5) \microk to (17.5$\pm$9.5) \mmicrok, we 
account for the residual dust contribution.
\label{tab:tabff}
\end{table}

%%%%%%%%%%%%%%%%%%%%%% REFERENCES: %%%%%%%%%%%%%%%%%%%%%%%%%

\clearpage

\setcounter{secnumdepth}{0}

\section{REFERENCES}

 \rn Bennett, C.L., Smoot, G.F., Hinshaw, et al. \ 
%\rn Bennett, C.L., Smoot, G.F., Hinshaw, G., Wright, E.L., Kogut, A.,
% De Amici, G., Meyer, S.S., Weiss, R., Wilkinson, D.T., Gulkis, S., 
% Janssen, M.A., Boggess, N.W., Cheng, E.S., Hauser, M.G., Kelsall, T.,
% Mather, J.C., Moseley, S.H., Murdock, T.L., Silverberg, R.F. \
  1992, ApJ, 396, L7 
% Preliminary separation of galactic and cosmic microwave emission for the
% COBE Differential Microwave Radiometer.

 \rn Bennett, C.L., Banday, A.J., Gorski, K.M., et al. \
%\rn Bennett, C.L., Banday, A.J., Gorski, K.M., Hinshaw, G., 
% Jackson, P., Keegstra, P., Kogut, A., Smoot, G.F., Wilkinson, 
% D.T., Wright, E.L. 
  1996, ApJ, 464, L1
% Four-Year COBE DMR Cosmic Microwave Background Observations: 
% Maps and Basic Results.

 \rn Boggess, N.W., Mather, J.C., Weiss, et al. \
%\rn Boggess, N.W., Mather, J.C., Weiss, R., Bennett, C.L., Cheng, E.S.,
% Dwek, E., Gulkis, S., Hauser, M.G., Janssen, M.A., Kelsall, T.,
% Meyer, S.S., Moseley, S.H., Murdock, T.L., Shafer, R.A., 
% Silverberg, R.F., Smoot, G.F., Wilkinson, D.T., Wright, E.L. \
  1992, ApJ, 397, 420 
% The COBE mission: its design and performance two years after launch.

 \rn Brandt, W.N., Lawrence, C.R., Readhead, A.C.S., et al. \
%\rn Brandt, W.N., Lawrence, C.R., Readhead, A.C.S., Pakianathan, J.N.,
% Fiola, T.M. \
  1994, ApJ, 424, 1
% Separation of foreground radiation from Cosmic Microwave Background
% anisotropy using multifrequency measurements.  
  
 \rn Fomalont, E.B., Partridge, R.B., Lowenthal, J.D., et al. \
%\rn Fomalont, E.B., Partridge, R.B., Lowenthal, J.D., Windhorst, R.A. \
  1993, AJ, 404, 8 
% Limits to cosmic background radiation fluctuations at 8.44 GHz between
% angular scales 10" and 200".

 \rn Gaustad, J.E., McCullough, P.R., and Van Buren, D.\ 
  1996, PASP 108, 351 
% An upper limit on the contribution of galactic free-free emission
% to the cosmic microwave background near the north celestial pole

% rn Gautier, T.N., III, Boulanger, F., Perault, M., et al. \
%\rn Gautier, T.N., III, Boulanger, F., Perault, M., Puget, J.L. \
% 1992, AJ, 103, 1313 
% A calculation of confusion noise due to infrared cirrus.
% Astronomical Journal, April 1992, vol.103, (no.4):1313-24.

 \rn Haslam, C.G.T., Klein, U., Salter, C.J., et al. \
%\rn Haslam, C.G.T., Klein, U., Salter, C.J., Stoffel, H.,
% Wilson, W.E., Cleary, M.N., Cooke, D.J., Thomasson, P. \
  1981, A\&A, 100, 209 
% A 408 MHz all-sky continuum survey. I. Observations at southern
% declinations and for the north polar region.

 \rn Kogut, A., Banday, A.J., Bennett, C.L., et al. \
%\rn Kogut, A., Banday, A.J., Bennett, C.L., Gorski, K.M., 
% Hinshaw, G., Reach, W.T. \
  1996a, ApJ, 460, 1 (K96a) 
% High-latitude Galactic emission in the COBE differential microwave
% radiometer 2 year sky maps.

 \rn Kogut, A., Banday, A.J., Bennett, C.L., et al. \
%\rn Kogut, A., Banday, A.J., Bennett, C.L., Gorski, K.M., 
% Hinshaw, G., Smoot, G.F., Wright, E.I. \
  1996b, ApJ, 464, L5 (K96b) 
% Microwave emission at high Galactic latitudes in the four-year 
% DMR sky maps.

 \rn K\"uhr, H., Witzel, A., Pauliny-Toth, I.I.K., et al. \
%\rn K\"uhr, H., Witzel, A., Pauliny-Toth, I.I.K., Nauber, U. \
  1981, A\&AS, 45, 367 
% A catalogue of extragalactic radio sources having flux densities 
% greater than 1 Jy at 5 GHz.

 \rn Netterfield, C.B., Devlin, M.J., Jarosik, N., et al. \
%\rn Netterfield, C.B., Devlin, M.J., Jarosik, N., 
% Page, L., Wollack, E.J. \ 
  1997, ApJ, 474, 47 (N97)
% A Measurement of the Angular Power Spectrum of the 
% Anisotropy in the Cosmic Microwave Background 

 \rn Partridge, R.B. \ 
  1995, 3K:The CMBR (GB: Cambridge University)
% 3K:The Cosmic Microwave Background Radiation. (GB: Cambridge University)

%\rn Reach, W.T., Dwek, E., Fixsen, D.J., et al. \
% 1995, ApJ, 451, 188 
% Far-infrared spectral observations of the Galaxy by COBE.

 \rn Reich, P., Reich, W. \ 
  1988, A\&AS, 74, 7 
% A map of spectral indices of the Galactic radio continuum emission
% between 408 MHz and 1420 MHz for the entire northern sky. 

 \rn Reynolds, R.J. \ 
  1992, ApJ, 392, L35 
% The optical emission-line background and accompanying emissions at
% ultraviolet, infrared, and millimeter wavelengths.

 \rn Simonetti, J.H., Dennison, B., Topasna, G.A. \ 
 1996, ApJ, 458, L1 
% The contribution of Galactic free-free emission to anisotropies in the
% cosmic microwave background found by the Saskatoon experiment.

 \rn Tegmark, M., Efstathiou, G. \ 
  1996, MNRAS, 281, 1297 
% A method for subtracting foregrounds from multifrequency CMB sky
% maps.

 \rn Tegmark, M., de Oliveira-Costa, A., Devlin, M.J., et al.\
%\rn Tegmark, M., de Oliveira-Costa, A., Devlin, M.J., 
% Netterfield, C.B., Page, L., Wollack, E.J. \ 
  1997, ApJ, 474, L77 (T97) 
% A High-Resolution Map of the Cosmic Microwave Background around the 
% North Celestial Pole

 \rn Wollack, E.J., Devlin, M.J., Jarosik, N., et al. \
%\rn Wollack, E.J., Devlin, M.J., Jarosik, N., 
% Netterfield, C.B., Page, L., Wilkinson, D. \ 
  1997, ApJ, 476, 440 (W97)
% An Instrument For Investigation of the Cosmic Microwave 
% Background Radiation at Intermediate Angular Scales 

%%%%%%%%%%%%%%%% FIGURE CAPTIONS: %%%%%%%%%%%%%%%%%%%%%%%%%

\clearpage 

\section{FIGURE CAPTIONS:}

\setcounter{figure}{0}

\begin{figure}[phbt]
\caption{ 
Saskatoon and template maps.
For all maps, the temperatures are shown in coordinates where the
North Celestial Pole is at the center of a circle of 15$^{\circ}$
diameter, with RA=0 at the top and increasing clockwise.
\From left to right, the three top panels show the 
Saskatoon Ka-Band map (Ka), the Saskatoon Q-Band map (Q) and the 
full Saskatoon (Ka+Q Bands) map (All). 
The three middle panels show the 408~MHz Haslam survey (Has), 1420~MHz 
Reich \& Reich survey (RR) and the point source template (PS).
Finally, the three bottom panels show the DIRBE 240, 140 and
100 \microm maps.
}
\label{maps}
\end{figure}

\begin{figure}[phbt]
\caption{ 
Cumulative probability distribution for the correlation 
between DIRBE 100 \microm and Saskatoon Q-Band data. The solid curve is 
for the case where $\alphaHat$ 
is Gaussian with zero mean and variance given by equation (\ref{varalpha}). 
The dotted curve is for the 288 different DIRBE sky patches. The horizontal
lines indicate the confidence levels of 68\% and 95\%, while the 
vertical line at $\alphaHat$=15.0 represents the correlation value given
by DIRBE template at the Saskatoon observing region.
}
\label{histograms}
\end{figure}

%%%%%%%%%%%%%%%% FIGURES: %%%%%%%%%%%%%%%%%%%%%%%%%%%%%%%%%%

% \clearpage

% \begin{figure}
% \begin{center}
% \centerline{\rotate[r]{\vbox{\epsfxsize=15cm\epsfbox{../../dirbe/idl/ducksmall.ps}}}}
% %\vspace{-1.cm}
% \caption{ 
% Saskatoon and template maps.
% For all maps, the temperatures are shown in coordinates where the
% North Celestial Pole is at the center of a circle of 15$^{\circ}$
% diameter, with RA=0 at the top and increasing clockwise.
% \From left to right, the three top panels show the 
% Saskatoon Ka-Band map (Ka), the Saskatoon Q-Band map (Q) and the 
% full Saskatoon (Ka+Q Bands) map (All). 
% The three middle panels show the 408~MHz Haslam survey (Has), 1420~MHz 
% Reich \& Reich survey (RR) and the point source template (PS).
% Finally, the three bottom panels show the DIRBE 240, 140 and
% 100 \microm maps.
% }
% \end{center}
% \label{maps}
% \end{figure}

% \clearpage

% \begin{figure}
% \begin{center}
% \centerline{{\epsfxsize=18cm\epsfbox{../../dirbe/idl/histogram.ps}}}
% %\vspace{-2.cm}
% \caption{ 
% Cumulative probability distribution for the correlation 
% between DIRBE 100 \microm and Saskatoon Q-Band data. The solid curve is 
% for the case where $\alphaHat$ 
% is Gaussian with zero mean and variance given by equation (\ref{varalpha}). 
% The dotted curve is for the 288 different DIRBE sky patches. The horizontal
% lines indicate the confidence levels of 68\% and 95\%, while the 
% vertical line at $\alphaHat$=15.0 represents the correlation value given
% by DIRBE template at the Saskatoon observing region.
% }
% \end{center}
% \label{histograms}
% \end{figure}

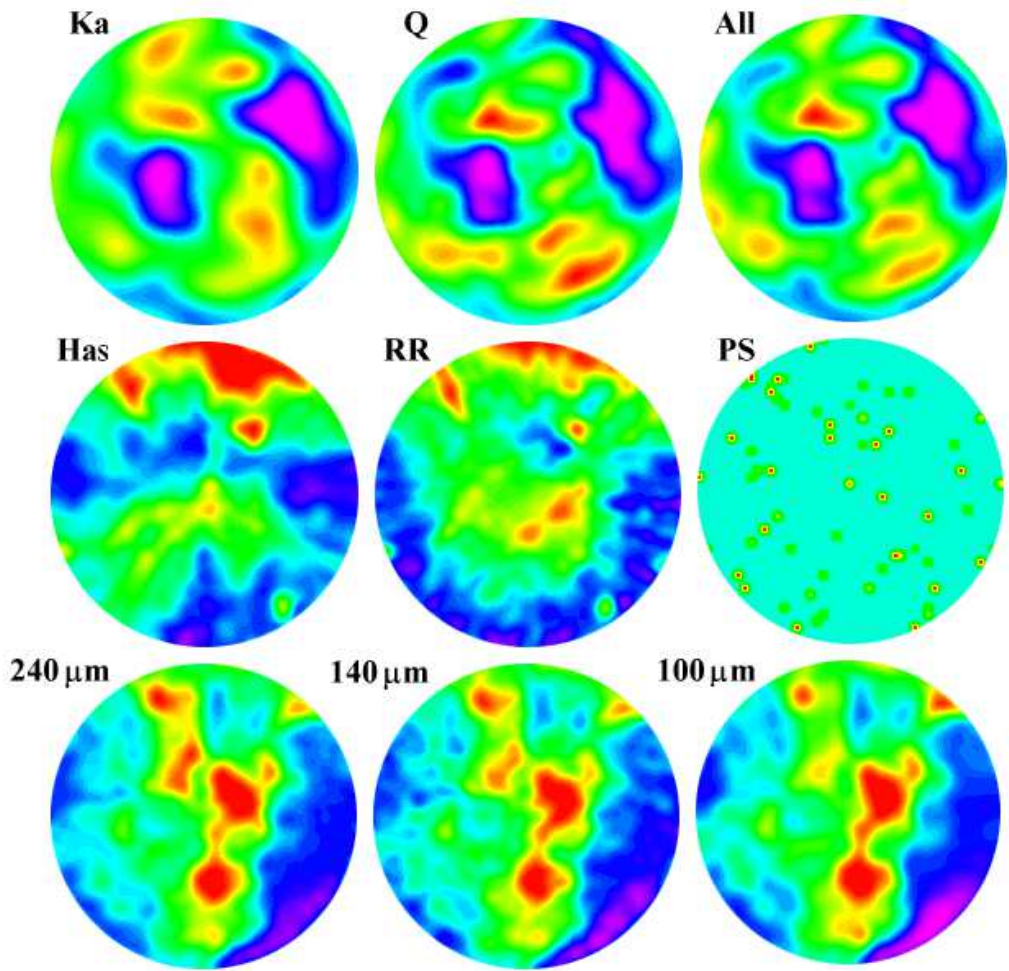
\end{document}